\documentclass[%
 aps,
 prx,%
 showpacs,
 amsmath,amssymb,floatfix,superscriptaddress,
 reprint,%
]{revtex4-2}
\usepackage{nameref}
\usepackage{soul}
\usepackage{hyperref}
\usepackage{graphicx}% Include figure files
\usepackage{bm}% bold math
\usepackage{color}
\usepackage[bf]{subfigure}
\begin{document}

\title{Unfolding of Crumpled Thin Sheets}   

\author{Leal F. C. B.}

\email{fbonleal@gmail.com}
\affiliation{Departamento de Física, Universidade Federal de Pernambuco, Recife, PE 50670-901, Brazil\looseness=-1}
\author{Gomes M. A. F.}
\email{mafg@ufpe.br}
\affiliation{Departamento de Física, Universidade Federal de Pernambuco, Recife, PE 50670-901, Brazil\looseness=-1}

\date{\today}
\begin{abstract}

Crumpled thin sheets are complex fractal structures whose physical properties are influenced by a hierarchy of ridges. In this Letter, we report experiments that measure the stress-strain relation and show the coexistence of phases in the stretching of crumpled surfaces. The pull stress showed a change from a linear Hookean regime to a sublinear scaling with an exponent of $0.65 \pm 0.03$, which is identified with the Hurst exponent of the crumpled sheets. The stress fluctuations are studied, the statistical distribution of force peaks is analyzed and it is shown how the unpacking of crumpled sheets is guided by long distance interactions.

\end{abstract}
%\begin{document}

%\pacs{05.40.-a, 05.70.Np, 68.35.Gy, 87.15.Cc}

\maketitle

\newpage

%\section{Introduction}

The unpacking of structures with many folds is a process of immense importance, as seen in the unpacking of nucleic acids \cite{DNA1,DNA2}, proteins \cite{deGennes} and other biological structures \cite{besouro}, such as the flowering of a flower bud, to name just a few cases that occur with an absurdly high frequency in nature. Here, processes of this type are studied in detail for the first time when considering the unpacking of crumpled surfaces. In addition to the intrinsic interest in the physical and mathematical aspects of unpacking processes, these are of great importance as they are the reverse operations of the crumpling mechanisms which, in turn, have been considered in many high-tech devices and structures in the field of electronics and materials science \cite{crump_batter,crump_capacitor,crump_meta}.

\begin{figure}[htp]

    \includegraphics[width=0.75\linewidth]{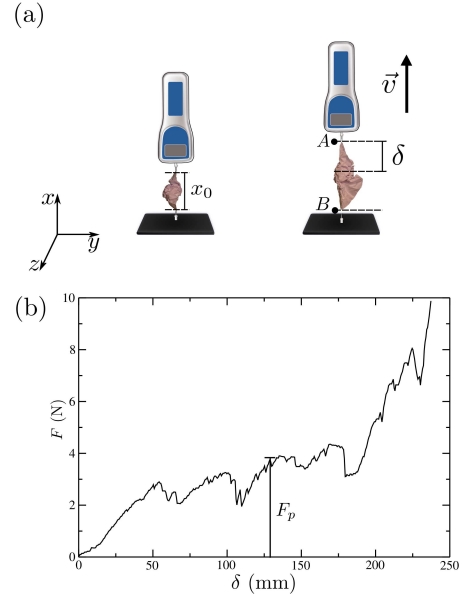}
    \caption{(a) Scheme of equipment used to stretch the paper balls. As the dynamometer is displaced with constant velocity $\vec{v}$ upwards the deformation  ``$\delta$'' increases. (b) The stretching force $F$ versus $\delta$ for a typical paper ball. See second to fourth paragraphs in the text for details.
    }
    \label{fig:1}
\end{figure}

This Letter reports fully automated experiments of unidirectional unfolding of near spherical crumpled paper balls made of square sheets of area $L^2$ \cite{Gomes1987}, under controlled ambient conditions, in order to obtain the curve unpacking force versus deformation. The stretching speed, $|\Vec{v}|$, was constant and the deformation $\delta$, measured along the pull axis $x$, is the full opening $AB$ minus the initial distance $x_0$ as shown in Fig.~\ref{fig:1}-a.

The unpacking of crumpled surfaces was performed for balls made with sheets of density $\mu=75$ $g/m^2$ and sizes $L = \{30, 55, 66, 77, 88, 147, 206, 264$ and $305$ $mm \}$. We reproduced the experiment 10 times for each value of $L$, with the stretching speed $v=0.83 \pm 0.01$ $mm/s$. Additional tests, with speeds between $0.83$ and $6.10$ $mm/s$ showed no significant variation in stress-strain curves.

%\section{Results}
%\subsection{Analise da curva de força}

A typical curve of force versus the stretching deformation, $F(\delta)$, of a paper ball is shown in Fig.~\ref{fig:1}-b. Fluctuations around the force curve occur due the unfolding of various types of local tangled folds, a phenomenon that has been described in detail for the strain-strain curve associated with the unfolding of crumpled thin sheet \cite{Leal2019}. Each peak signals that the system has moved from one local equilibrium state to another. This is a characteristic found in many types of physical systems exhibiting transition between metastable states \cite{metaR1,metaR2,metaR3,metaR4}. A similar phenomenon has been described in the unpacking of crumpled wires in two-dimensional cavities \cite{unpacking}. A relationship between the extraction force and the length of extracted wire can also be found in that work.

\begin{figure*}[t]
\centering\includegraphics[width=0.9\textwidth]{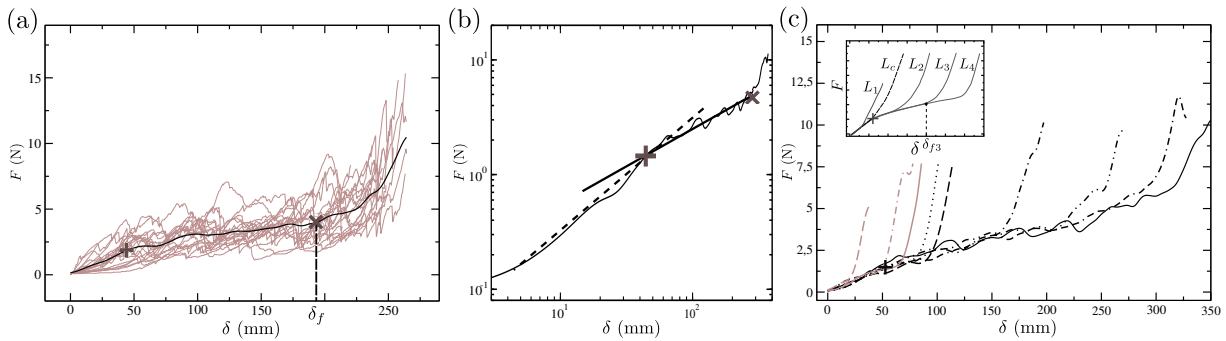}
\caption{(a) The sepia curves are 20 sequences of measurements of force, $F$, versus stretching, $\delta$, equivalent to Fig.~\ref{fig:1}-b. To reduce fluctuations in the 20 sequences we use the Nadaraya-Watson method, which can be seen in the continuous averaged curve. (b) Log-log graph of $F$ versus $\delta$ Nadaraya-Watson curve for a sheet with $L = 305$ mm. The dashed line shows a $F \sim \delta^{1.04 \pm 0.01}$ power law scaling. The continuous line shows a change in the $F$-$\delta$ curve in the second region, to the power law $F \sim \delta^{0.65 \pm 0.03}$. (c) Nine Nadaraya-Watson curves averaged on 10 equivalent sequences $F$ versus $\delta$, for $L = \{33,55,66,77,88,147,206,264$ and $305$ $mm \}$, (for more details see the text). In the inset we have the schematic behavior of the average curves of $F$-$\delta$ for different $L$.}\label{fig:2}
\end{figure*}

In Fig.~\ref{fig:2}-a, we have a graph with 20 equivalent stretching experiments with the continuous curve representing the average force curve generated with the non-parametric regression method of Nadaraya and Watson \cite{Nadaraya1964,Watson1964} whose purpose is to reduce fluctuations. In this same graph, we can clearly see that the continuous $F$-$\delta$ curve has three distinct trends. Due to their importance we have placed symbols to represent the two points of transition between them. The cross ($+$) represents the point of change from the first region to the second ($\delta_{(+)}$ is the abscissa value of the point “$+$”) and the “$\times$” the point at which the second region ends ($\delta_{f}$ is the abscissa value of the point “$\times$”). $\delta_{f}$ indicates the beginning of the region, at the end of the $F(\delta)$ curve, where $F$ starts to grow very quickly, that is, it indicates the region where the internal cohesion forces internal to the sheet start to control the physics of the process. Thus, $\delta_{f}$ is associated with the transition from crumpled surface (CS) physics in three dimensions to physics controlled by the two-dimensional topology of the sheet. In Supplementary Material A (Figure S1) the reader can see images of the CS unfolding and their respective $\delta$.

%Therefore, we should expect that $\delta_{f} \sim L$, where L is the relevant topological length.

%\subsection{Características da curva F-Dx}

The trend of the average curves, described in Fig.~\ref{fig:2}-a, in the first and second regions obey two different scaling laws. We see this in the log-log graph of Fig.~\ref{fig:2}-b which shows a Nadaraya-Watson curve of $F$-$\delta$ for a sheet with $L = 305$ mm. The average curve obeys a double power law dependence

\begin{equation}\label{eq:F_Dx}
    F \sim \delta^n,
\end{equation}
where $n$ is $1.04 \pm 0.01$ for $\delta < \delta_{(+)}$, that is a linear Hookean behavior in the first region, and $0.65 \pm 0.03$ for $\delta_{f} > \delta > \delta_{(+)}$. This sublinear behavior found here, typical of auxetic materials, makes such materials exhibit the important feature of mitigating impacts \cite{auxetico_stress_train}.

Typical stress-strain curves in usual non-fragile materials such as metals and polymers, initially exhibit a linear behavior, characteristic of a Hookean regime, followed by a nonlinear, reversible regime, and then an irreversible plastic regime. In all materials, the behavior of the curve in the plastic regime varies a lot depending on the composition of the material, however, it usually has a ``convex shape'' and generally the curve ends at the breaking point \cite{metal_Stress_S, polim_Stress_strain}. Differently, the stress-strain curve of CS stretching has a ``concave shape'', between the second and third regions in Fig.~\ref{fig:2}-a, due to the underlying structure that is controlled by the 2D topology of the sheet. The same aspect can be seen in stress-strain curves of stretching experiments with individual molecules of DNA or proteins until they are fully unfolded (or unpacked) \cite{DNA1,DNA2}, which are also influenced by an underlying topology of reduced dimension (1D) as compared to that of the embedding three-dimensional space.

By varying the scale $L$ of the system we were able to observe other characteristics related to the slopes of the $F$-$\delta$ curve. The graph in Fig.~\ref{fig:2}-c shows nine unfolding curves of crumpled sheets that reflect the Nadaraya-Watson method for 10 equivalent experiments. To facilitate the reader's understanding, in the Figure~\ref{fig:2}-c inset we have a scheme simplifying the behavior of the averages curves of $F$-$\delta$. The slope of the curves in the first region is invariant by scale, that is, it does not change with $L$. The average curves shown in Fig.~\ref{fig:2}-c are associated with two unpacking behaviors. In the first behavior, the mean curves have only two regions, the first and the third, differently from the mean curve in Fig.~\ref{fig:2}-a (black continuous curve), which have three different regions. These curves have $L \leq L_{c}$, where $L_{c}$ is the size that marks the transition between the two behaviors (the dashed curve in the inset corresponds to the sheet with size $L_c$). These curves lack the sublinear intermediate region, this means that they pass continuously from the first to the third region. Because of this, their corresponding values of $\delta_{f}$’s are less than $\delta_{(+)}$. Note that the point of change from the first region to the second ($+$) is the same for all these curves. The sheet with size $L_c$ does not have the second region and its $\delta_{f}$ is equal to $\delta_{(+)}$. The experiment represented in the graph of Fig.~\ref{fig:2}-c that comes closest to $L_c$ corresponds to $L = 66 \pm 1$ mm with $\delta_{f} = 58 \pm 12$.

For the second behavior, the curves of the averages $F$-$\delta$ represent CSs with $L > L_{c}$, and have three different regions similar to Fig.~\ref{fig:2}-a. The same behavior is exemplified in the $L_3$ and $L_4$ curves in the inset of Fig.~\ref{fig:2}-c; there the stress-strain plots follow the same path in the graph $F$-$\delta$ and separate at the point with abscissa $\delta_{f3}$. This means that the work to unpack the ball $L_3$, $W_{3}(\delta_{f3})$, is equal to the work to stretch the ball $L_4$, $W_{4}(\delta_{f3})$, up to the point of abscissa $\delta_{f3}$. So $W_{i}(\delta_{fi}) = W_{j}(\delta_{fi})$, where $i < j$. Consequently

\begin{equation}
\label{eq:trab}
   W_{i}(\delta) = W_{j}(\delta),
\end{equation}
since $\delta \leq \delta_{fi}$. From detailed measurements of the work $W(\delta_{f})$ done to fully unpack the sheets in Fig.~\ref{fig:2}-c [See Supplementary Material B] it can be shown that both eq.~\ref{eq:trab} and the sublinear power law found in eq.~\ref{eq:F_Dx}, for $\delta_{f}> \delta > \delta_{(+)}$, are consistent equations of the same dynamical system.

\begin{figure*}[t]
\centering\includegraphics[width=0.9\textwidth]{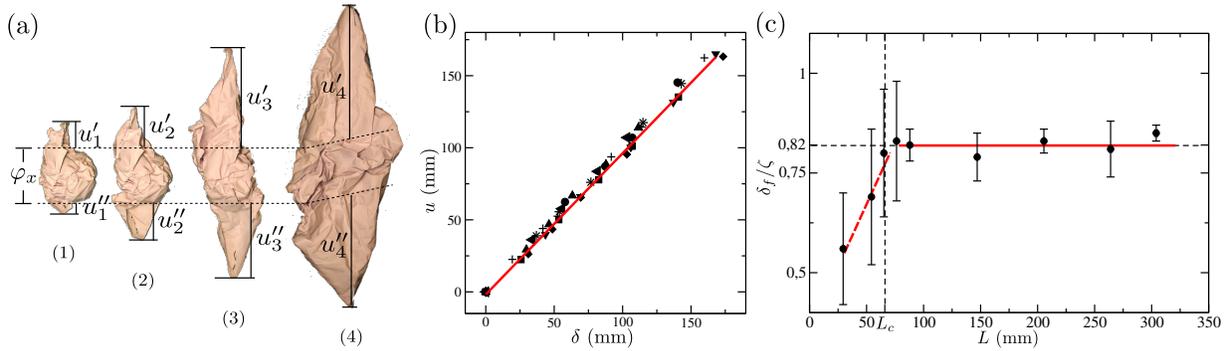}
\caption{(a) Sequence of images showing the evolution of the stretching process, in order to find the total length $u$ of the unpacked phase, $u_{i} = u'_{i} + u''_{i}$ as a function of $\delta$. The diameter of the packaged phase in the direction $x$, $\varphi_{x}$, is constant. (b) The graph of $u$ versus $\delta$ shows a linear fit $u = (1.01 \pm 0.01) \delta$. (c) Graph $\delta_{f}/\zeta$ versus $L$, where the maximum stretch ($\zeta$) is equal to $\sqrt{2} L - x_{0}$.}\label{fig:3}
\end{figure*}

Now, we examine how the packed and the unpacked phases evolve by measuring a characteristic length $u$ of the unpacked phase. In the Fig.~\ref{fig:3}-a we have a sequence of images that shows the evolution of the stretching process of a CS of initial radius $R = 18 \pm 1$ mm made with a sheet of $L = 212$ mm. The four images reveal that the unfolding process propagate from the clampled extremities to the middle of the sheet with an up-down symmetry along the strain direction. We measure the total length characterizing the degree of unfolding of the unpacked phase, $u_{i} = u'_{i} + u''_{i}$, where $u'_i$ and $u''_{i}$ are the displacements shown in Fig.~\ref{fig:3}-a. The graph in Fig.~\ref{fig:3}-b shows the dependence of $u$ versus $\delta$ for ten crumpled sheets (on each such sheets was performed between 16 to 20 measurements), that takes us to the equation,

\begin{equation}
\label{eq:Du_Dx}
    u = \delta.
\end{equation}

The unpacked part of the sheet corresponds to an area $A_{u}$ that obeys a relationship similar to the eq.~\ref{eq:trab} (See Supplementary Material C for more detail), that is,

\begin{equation}
\label{eq:Af}
    A_{ui}(\delta) = A_{uj}(\delta)
\end{equation}
where the $A_{ui}$ and $A_{uj}$ indexes indicate the equal unpacked areas of two CS with different $L$ sizes. Equations \ref{eq:trab} and \ref{eq:Af} mean that the mechanisms responsible for the cohesive forces of the paper ball are independent of the parameter $L$. The results presented in eqs.~\ref{eq:Du_Dx} and \ref{eq:Af} are not obvious, because the CS has a heterogeneous structure with long-range correlations that make unpacking a complex process (more on this ahead). As shown in Fig.~\ref{fig:3}-a the extension of the packaged part of the CS in the direction $x$ does not change with $\delta$, so the packaged phase has a frozen structure that is maintained as far as ``$\delta_{f}$'' is not reached. Fig.~\ref{fig:3}-c shows the graph of $\delta_{f}/\zeta$ vs. $L$, where $\zeta$ is the maximum possible stretch of a crumpled sheet, $\zeta = \sqrt{2} L - x_{0}$. Curves with $ L > L_{c}$ have $\delta_{f}/\zeta \approx 0.82 \pm 0.02$, that is, $\delta_{f}$ is proportional to $L$, since $\zeta$ is proportional to $L$. CS are fractal structures with complex interactions, so we must expect that the proportionality between $\delta_{f}$ and $L$ has a non-trivial reason. In Supplementary Material D we show that the CS stretch has a symmetry of unfolding of the packaged phase that is related to the proportionality relationship between $\delta_{f}$ and $L$. On the other hand, for sheets with $L \leq L_{c}$, $\delta_{f}/\zeta$ grows as $\delta_{f}/\zeta = (0.36 \pm 0.06) + (0.007 \pm 0.001) L$, that is $\delta_{f}$ depends quadratically on $L$, meaning a different unfolding mechanism. The value found for $L_c$ by adjusting the curve $\delta_{f}/\zeta $ versus $L$ in Fig.~\ref{fig:3}-c is $66 \pm 9$ mm and $\delta_{+} = 54 \pm 7$ mm.

%O diâmetro da fase empacotada na direção $x$, $\varphi_{x}$, é constante.

%*************************** Modelo *******************************

\begin{figure}[!ht]
    \includegraphics[width=0.8\linewidth]{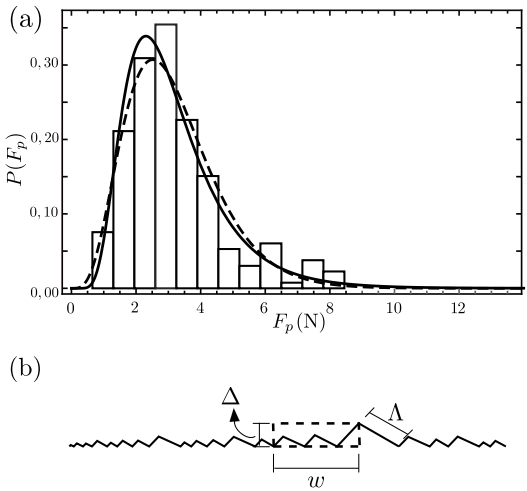}
    \caption{(a) Averaged peak size distribution $P(F_p)$ for ten equivalent samples of crumpled sheets with $L = 264$ mm. (b) Diagram of the cross section of a small local portion of a crumpled sheet, showing the roughness $\Delta$ for a range $w$ and the length $\Lambda$ of the facets.}
    \label{fig:dist_Fp}
\end{figure}

The mechanics underlying the whole phenomenology of stretching the crumpled sheet of paper has its roots in the organization of the ridges of the sheet. The most basic elements of the $F$-$\delta$ curve are the fluctuation peaks, which occur during sudden buckling changes in the facets that define the rough conformation, similar to what was described in the study of the noise emitted by a crumpled elastic sheet \cite{Kramer1996}. To find the relationship between the peaks of the $F$-$\delta$ curve and the length of the facets ($\Lambda$) we analyze the magnitude distribution of the peaks forces $F_p$. The magnitudes of $F_p$ correspond to the values of the force measured at each peak, as highlighted in Fig.~\ref{fig:1}-b. The distribution of $F_p$ for 10 equivalent curves $F$-$\delta$ is shown in Fig.~\ref{fig:dist_Fp}-a. The continuous line is a fit with the lognormal distribution which seems to capture quite well the characteristics of the data. Then, the magnitude of the peaks of force observed here and of the size of facets ($\Lambda$) reported in \cite{Blair2005} obey a similar hierarchical relationship controlled by a same statistical distribution. The data in Fig.~\ref{fig:dist_Fp}-a also presents a good fit for the Gamma distribution (dashed curve). For more details on the distribution of $F_p$ see Supplementary Material E.

It is important to observe how the length of the facets is distributed over the surface: the topography of the crumpled sheet as revealed by laser scanning \cite{Blair2005} shows that the average ridge roughness (or average elevation of the roughness) $\bar{\Delta}$, of a unfolded crumpled sheet of paper, within a box with range $w$ (see these quantities in Fig.~\ref{fig:dist_Fp}-b), scale as $\bar{\Delta} \sim w^H$. This power law is a Hurst analysis \cite{Blair2005} of the topography of the crumpled sheet, with an exponent of Hurst $H = 0.71$, for $w > w_{c}$, and $H = 1$, for $w < w_{c}$, where $w_{c} = 25 \pm 6$ mm \footnote{We estimate $\log_{10} {w_{c}} \approx 1.4 \pm 0.1$ from Fig. 5-c of ref. \cite{Blair2005}.}. As the facets are delimited by the network of ridges, then $\bar{\Delta} \propto \bar{\Lambda}$, thus

\begin{equation}
\label{eq:topg}
    \bar{\Lambda} \sim w^H.
\end{equation}
The parallel between $F_p$ and $\Lambda$ pointed at the end of the previous paragraph, and the matching between deformation ($\delta$) with the range ($w$) as the unfolding proceeds physically means that

\begin{equation}\label{eq:Fp}
    \bar{F}_{p} \sim \delta^H,
\end{equation}
where $H \approx 0.70$ for $\delta_{(+)} < \delta < \delta_{f}$ and $H \approx 1.0$ for $\delta < \delta_{(+)}$. Equation \ref{eq:Fp} is similar to  eq.~\ref{eq:F_Dx} and this leads us to identify the exponent of the first and second regions of the $F$-$\delta$ curve with the Hurst exponent $H$.

%, the factor $\frac{1}{2}$ because the CS unfolding occurs in two parts at the same time.

%The value $\delta' = 2 w_{c} = 2 (25 \pm 6) = 50 \pm 12$ is incredibly similar to the result found by the settings in Fig.~\ref{fig:3}-c, which got $\delta_{(+)} = 54 \pm 7$.

%%%%%%%%%%%%%%%%%%%%%%%%%%%%%%%%%%%%%%%%%%%%%%%%%%%%%%%%

Two works that examine thin sheet folding when a radial compression force  $f$ is applied showed that the sheets go through two regimes which are characterized by curvature patterns \cite{Vliegenthart2006,Tallinen2009}. In the first regime, the compression radius $\Phi$ of the confinement volume is greater than $\Phi_{c}$ and $f < f_c$, where $\Phi_{c}$ and $f_c$ are the transition radius and the force of transition to the crumpled regime. This initial regime of smaller curvatures (less accentuated than in a ridge), is characterized by having conics developable in elastic sheets, this means that the work of the force $f$ is stored in the form of elastic energy. The fractal dimension in this regime is similar to that valid for the uncrumpled sheet, $D = 2$. For $\Phi < \Phi_{c}$ and $f > f_c$, the sheet is in the crumpled regime, it has a hierarchy of ridges and $f$ obeys the scaling $f \sim (\kappa/L) (L/h)^{\beta} (\Phi/L)^{-(\alpha (\beta + 1)-1)}$, where $\kappa$ is the bending modulus and $h$ is the thickness of the sheet. The exponent $\alpha$ is found through simulation from \cite{Tallinen2009}:

\begin{equation}
\label{eq:fac}
   \bar{\Lambda} \approx L.(\Phi/L)^{\alpha},
\end{equation}
with $\alpha = 1.65$, and with $\beta = 1/3$ coming from the well-known scaling law for the energy of a single ridge \cite{Lobkovsky1995}. The sheet fractal dimension in this regime is $D = 2.5$ and $\Phi_{c}/L = 0.4$ \cite{Tallinen2009} for any size $L$ of an elasto-plastic sheet. After replacing $\Phi_{c}/L$ with $0.4$ in $f(h,\Phi,L)$ we get $f_{c}(h,L) = C(h).L^{(\beta - 1)}$, with $C(h)$ being a function $h$. In the graph of figure \ref{fig:2}-c we have a record of the stretching of sheets of different sizes that were compressed with the hand grip force $f_m$. Through the inverse function of $f_{c}(h,L)$ we find the sheet with size $L = L'$ satisfying $f_{c} = f_{m}$, that means that all sheets with $L > L'$ are in the crumpled regime and have a fractal dimension $D = 2.5$. The sheets with $L \leq L'$ are in the regime of developable conics and consequently fractal dimension equal to that of surfaces, $D = 2$. Assuming that $L_{c} = L'$, the change of regime found in the stretching of CS, between $L \leq L_c$ and $L > L_c$, is caused by the unfolding of the crumpled sheet from different structural regimes (regime of developable conics or the crumpling regime). The characteristic length $L_c$ is a function of the force $f_m$ and $h$. Using the relationship between mass ($M$) and the radius of the paper ball ($R$) \cite{Gomes1987} as a boundary condition, $R = \Phi \sim L^{2/D}$, from eq.~\ref{eq:fac} we have

\begin{equation}
\label{eq:lamb2}
   \bar{\Lambda} \sim L^{[(2/D-1).\alpha+1]} \rightarrow \bar{\Lambda} \sim L^{H},
\end{equation}
where $H = (2/D - 1)\alpha + 1$. For sheets with $L < L_{c} \rightarrow D=2$ and $H = 1$. For sheets with $L > L_{c} \rightarrow D=2.5$ and $H = 0.67$.

The correspondence between eqs.~\ref{eq:topg} and \ref{eq:lamb2} means that we can also find the transition between the applicable conical regime (or small curvature regime) and the crumpled regime by varying the scale $w$. Therefore, on a sheet with $L > L_{c}$ and scale $w < w_{c}$ we find the applicable conical regime. Since $\delta \sim w$ then $\delta_{(+)}$ is associated with $w_{c}$, therefore the linearity found at the beginning of the $F$-$\delta$ curve ($\delta < \delta_{(+)}$) is related to the small curvature folds of the applicable conical regime. In a scale $w > w_{c}$, we find the crumpled regime and for this reason the curve $F$-$\delta$ has a sublinear regime for $\delta_{(+)} < \delta < \delta_{f}$. The exponent $H = 0.65$ indicates that CS roughness is not purely random, but has memory for distances greater than the average ridge length \cite{expHurst}.

Each stretched piece of the originally crumpled surface is a continuous variety formed by the union of facets whose average length is give by eq.~\ref{eq:fac}. Thus, the number of facets $N \approx (L/ \Lambda)^{2} = (L / \Phi)^{2 \alpha}$. The hierarchy of ridges and $F_p$ are associated with the facet partitioning of the sheet \cite{Wood2002}, so the eq.~\ref{eq:lamb2} connects two properties of CS, partitioning and long distance memory, due to the exponent $H > 0.5$. In a three-dimensional perspective, the evolution of partitioning is a process of overlapping ridges that occurs in a hierarchical order. The unpacking of a CS must obey the reverse process of that overlap. The overlap of ridges creates a cohesive structure that frustrates the unfolding when it does not follow the order established by the hierarchy of ridges. For this reason, the diameter of the packaged phase in the direction $x$, $\varphi_{x}$, in Fig.~\ref{fig:3}-a, is constant, as the energy required to unfold the ridges that are outside hierarchical order diverges. The equality present in eq.~\ref{eq:Du_Dx} is a consequence of this immobile state of the packaged phase and shows that the work of the stretching force, $W(\delta)$, is used completely to unfold the ridges that are in the correct order. If $\varphi_{x}$ changes with $\delta$, then part of the work $W(\delta)$ would be used for this change and the stretching tension would involve unfolding rigdes from different hierarchies at the same time.

%For this reason, the unfolding rate $du / d \delta$ is constant and the packaged phase in Fig.~\ref{fig:3} remains frozen, i.e., the diameter of the packaged phase in the direction $x$ does not change with $\delta$, as long as the distance prohibit unfolding outside the hierarchical order.

In this Letter, we studied in detail for the first time the unfolding of crumpled systems, observing the force necessary to unpack and stretch these complex structures. The corresponding force curve presents different regimes as the sheet is unfolded: a linear Hookean regime at the beginning of the stretching, as a result of the crumpled surfaces elasticity, and a sublinear regime that has an exponent equal to $0.65 \pm 0.03$, identified with the Hurst exponent of the crumpled sheet which in turn is associated with the surface roughness \cite{Gomes1989, Blair2005}. We show that the behavior of the stress curve is guided by long distance correlations and that the dynamics of the unfolding of the ridges is coordinated by a hierarchical order. In the unfolding process reported here there is the coexistence of two structural phases: a crumpled frozen fractal phase (packaged phase) and an open solid two-dimensional rough phase of a surface with fixed connectivity (unpacked phase).

This work was supported by Coordena\c{c}\~ao de Aperfei\c{c}oamento de Pessoal de N\'ivel Superior (CAPES), Programa PROEX 534/2018, $\#$ $23038.003382/2018$-39 and Conselho Nacional de Desenvolvimento Cientifico e Tecnol\'ogico (CNPq). The authors thank Daniel F. Gomes for help in the instrumentation. FCBL thanks a fellowship from CNPq, and stimulating discussions with Tiago T. Saraiva.\\
\textbf{Author contributions} Conceived and designed the experiments: FCBL and MAFG. Performed the experiments: FCBL. Analyzed the data: FCBL and MAFG. Wrote the paper: FCBL and MAFG.

%\section{Acknowledgment}
 
%\textcolor{red}{[OTHER GRANTS HERE!]}

\bibliographystyle{ieeetr}

\bibliography{bibligrafia}

\end{document}

\title{Supplementary Material for\\
``Unfolding of Crumpled Thin Sheets''}  

\author{F. C. B. Leal}
\email{fbonleal@gmail.com}
\affiliation{Departamento de Física, Universidade Federal de Pernambuco, Recife, PE 50670-901, Brazil\looseness=-1}
\author{M. A. F. Gomes}
\email{mafg@ufpe.br}
\affiliation{Departamento de Física, Universidade Federal de Pernambuco, Recife, PE 50670-901, Brazil\looseness=-1}

\date{\today}
\maketitle
\newpage

\subsection{Images of the unfolding of crumpled surfaces}\label{ape:ima_desd}

\begin{figure}[!ht]
    \centering
    \includegraphics[width=0.95\linewidth]{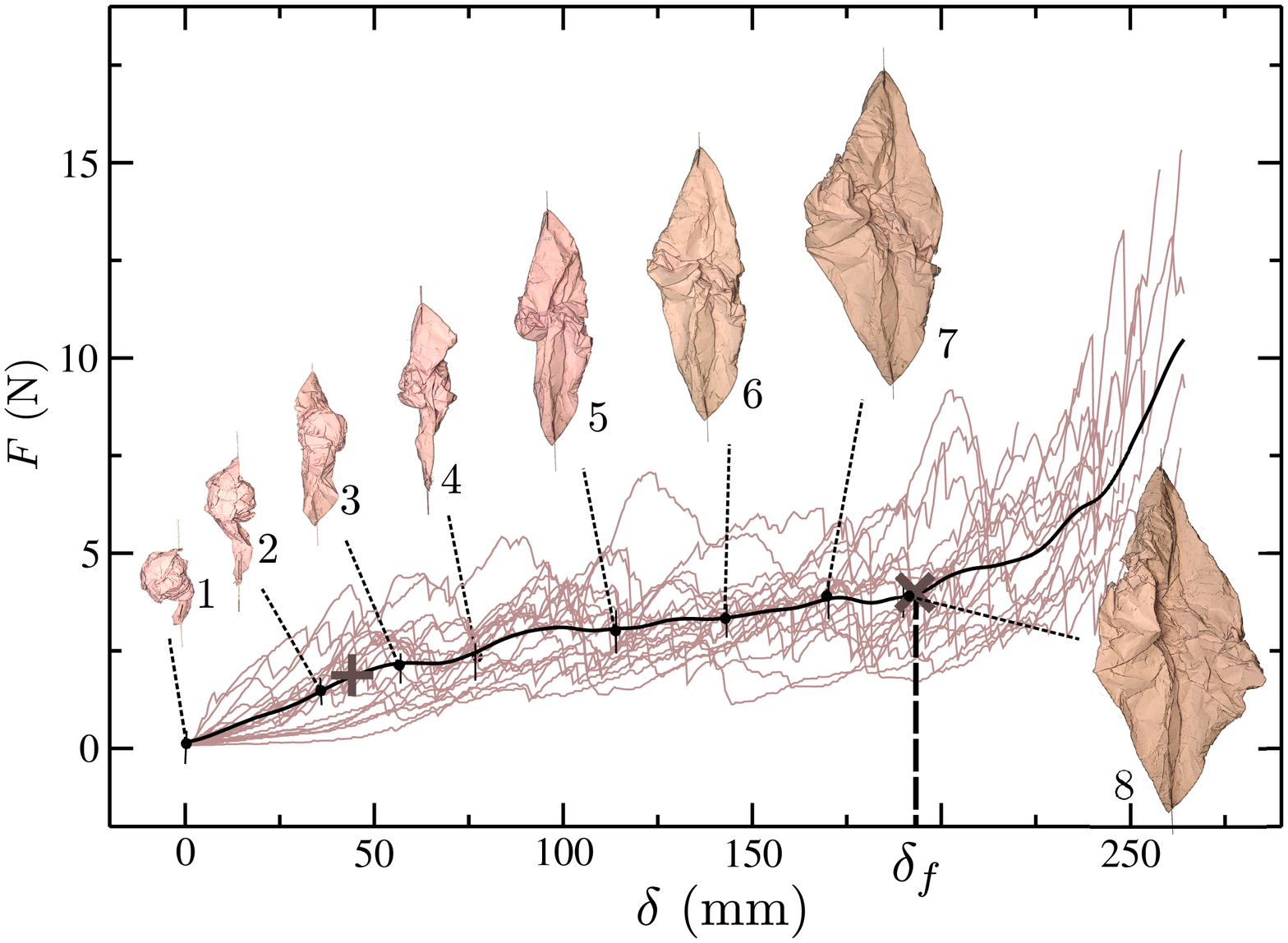}
    \caption{Experimental stress-strain $F$-$\delta$ curves obtained from 20 equivalent samples. The continuous black line refers to the Nadaraya-Watson method used to obtain reliable averages. The eight photographs show the evolution of CS unpacking.}
    \label{fig:foto_desd}
\end{figure}

The evolution of the unpacking of a crumpled paper ball is shown in Fig.~\ref{fig:foto_desd} in eight images taken from approximately equal time intervals. The uniaxial unfolding stress is directed along a diagonal of the sheet. Each image points to the region of the average curve where the respective deformation occurred. Although the images do not belong to any of the 20 crumpled surfaces (CS) sampled in the graph, they describe a typical unpacking of a crumpled sheet of 210x210 $mm^2$. Photo (1) refers to the axial pull strain $\delta = 0$ and it represents the state of maximum packing. Images (1) and (2) show two configurations of a CS in the first region.

The deformations recorded in the photos from (3) to (7) refer to the second region. These images show that two parts of the paper sheet progressively loosen themselves from a packaged phase of the sheet as $\delta$ grows, forming an unpacked rough phase that coexist with a more compact truly crumpled phase. These two phases coexist in the first and second regions of the $F$-$\delta$ curve shown in Fig.~\ref{fig:foto_desd}. However, at the point ($\times$) the packaged phase is completely dismantled as we can see from the photo (8). After this point, we have the third region where the surface is entirely in the unpacked phase, although the sheet is still rough. So, the steeper slope found in the third region is the result of a decrease in the roughness and, eventually, a stretching of the paper fibers. That is, in the first and second regions, the forces involved are locally transversal to domains of the sheet; they usually connect separate domains on the sheet. Differently, in the last region, the forces involved are basically internal to the sheet.

\subsection{The work for unpacking}\label{apen:trab}

\begin{figure}[!ht]
    \centering
    \includegraphics[width=0.8\linewidth]{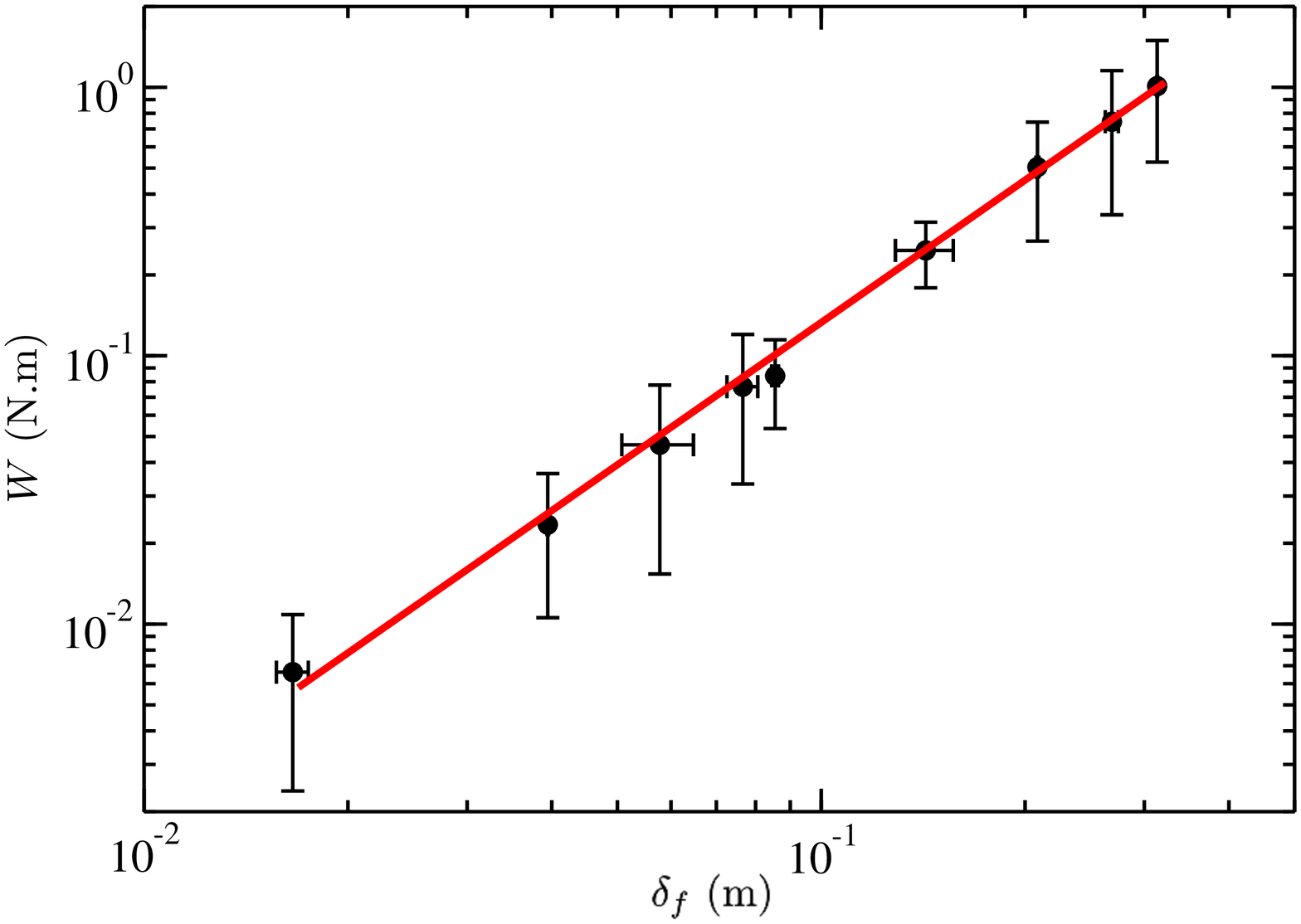}
    \caption{Work done, $W$, in SI units to unpack crumpled paper balls for different $L$ versus $\delta_{f}$. The data were obtained through the numerical integration of the curves $F$-$\delta$ of Fig. 2-c for nine different sizes $L$. The power law setting is $W = (6.9 \pm 0.1)\delta_{f}^{(1.73 \pm 0.05)}$.}
    \label{fig:w_Dx}
\end{figure}

Through the numerical integration of the nine $F$-$\delta$ curves of Fig. 2-c of the manuscript, we obtain the corresponding work $W(\delta_{f})$ to fully unpack the sheets as shown as by the graph in Fig.~\ref{fig:w_Dx}. The work obeys the power law

%Através da integração numérica das nove curvas $F$-$\delta$ da Fig. 2-c, nós obtemos suas medidas do trabalho ($W(\delta_{f})$) para desempacotar totalmente as folhas. Associando os $W(\delta_{f})$'s com os seus respectivos $\delta_{f}$’s, conforme sugerido pelo o gráfico da Fig.~\ref{fig:w_Dx}, nós obtemos um ajuste de lei de potência nas medidas $W$ vs $\delta_{f}$ que somada a eq. 3 do texto principal encontramos a relação

\begin{equation}
\label{eq:trabMS}\tag{S1}
   W(\delta) \sim \delta^{(1,73 \pm 0,05)}. 
\end{equation}
Deriving equation \ref{eq:trabMS} in relation to $\delta$ we get $F \sim \delta^{(0.73 \pm 0.05)}$, which is a reasonable approximation for the power law fit found in the second region of the $F$-$\delta$ curve of Fig. 2-b which presents an exponent $n = 0.65 \pm 0.03$.

%Derivando a equação \ref{eq:trabMS} em relação a $\delta$ temos $F \sim \delta^{(0,73 \pm 0,05)}$, que é uma boa aproximação para o ajuste de lei de potência encontrado na segunda região da curva $F$-$\delta$ da Fig.~2-b (onde obtemos um expoente $n = 0,63 \pm 0,05$).

\subsection{Detailing the coexistence of phases - I}\label{apend:coex_fase}

\begin{figure}[!ht]
    \centering
    \includegraphics[width=0.8\linewidth]{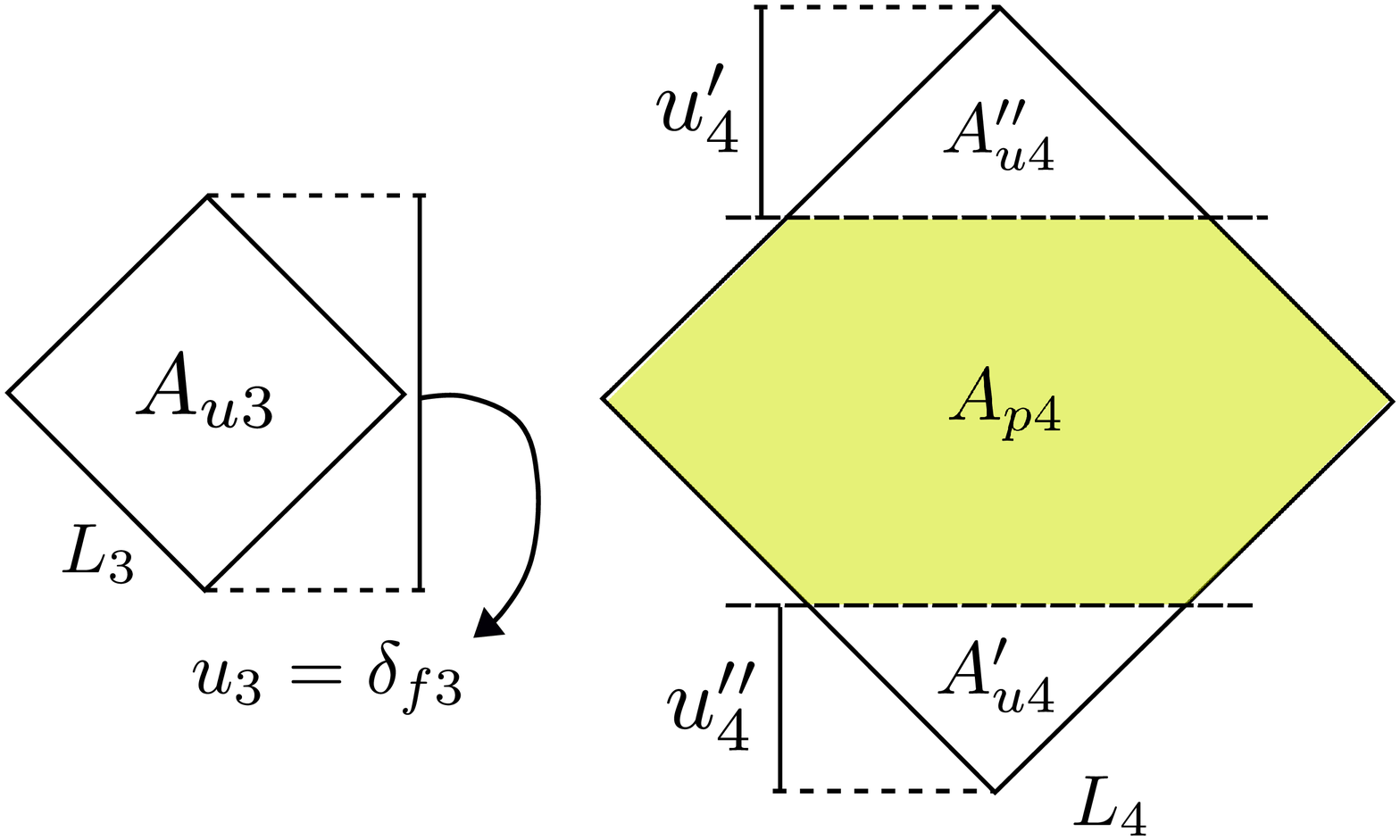}
    \caption{Map showing the areas of the sheets that are unpacked, $A_{u3}$ and $A_{u4}$, and the part that is still packed, $A_{p4}$. They have been stretched to $\delta_{f3}$, at this point the sheet $L_3$ is fully unpacked and the sheet $L_4$ is partially unpacked. The unpacked area of the two sheets are the same, $A_{u3} = A_{u4}$ because $\delta_{f3} = u_{3} = u'_{4} + u''_{4} = u_{4}$.}
    \label{fig:desd_fol}
\end{figure}

In Fig.~\ref{fig:desd_fol} we have two sheets with sizes $L_3$ and $L_4$ equally stretched up to $\delta = \delta_{f3}$, that is, to the point at which the sheet $L_3$ is completely unpacked and $A_{u3}$ is the unpacked area of the sheet. They have been mapped to show the parts of the sheets that are unpacked, $A_{u3}$ and $A_{u4}$, and the part that is still packed, $A_{p4}$. The sheet $L_4$ has the unpacked areas, $A'_{u4}$ and $A''_{u4}$, and the packed area, $A_{p4}$. Through eq. 3 of main text we arrive at the expressions $u_{3} = \delta_{f3}$ e $u'_{4} + u''_{4} = u_{4} = \delta_{f3}$ (since the sheet with size $L_4$ has been stretched to $\delta_{f3}$) so the unpacked areas of the two sheets are the same. Therefore, if two paper balls $i \neq j$ with different $L$'s are stretched to the same $\delta$, the average unpacked areas, $A_u$, of both will be the same,

\begin{equation}
\label{eq:area}\tag{S2}
   A_{ui}(\delta) = A_{uj}(\delta).
\end{equation}

\subsection{Detailing the coexistence of phases - II}

\begin{figure}[!ht]
    \centering
    \includegraphics[width=0.8\linewidth]{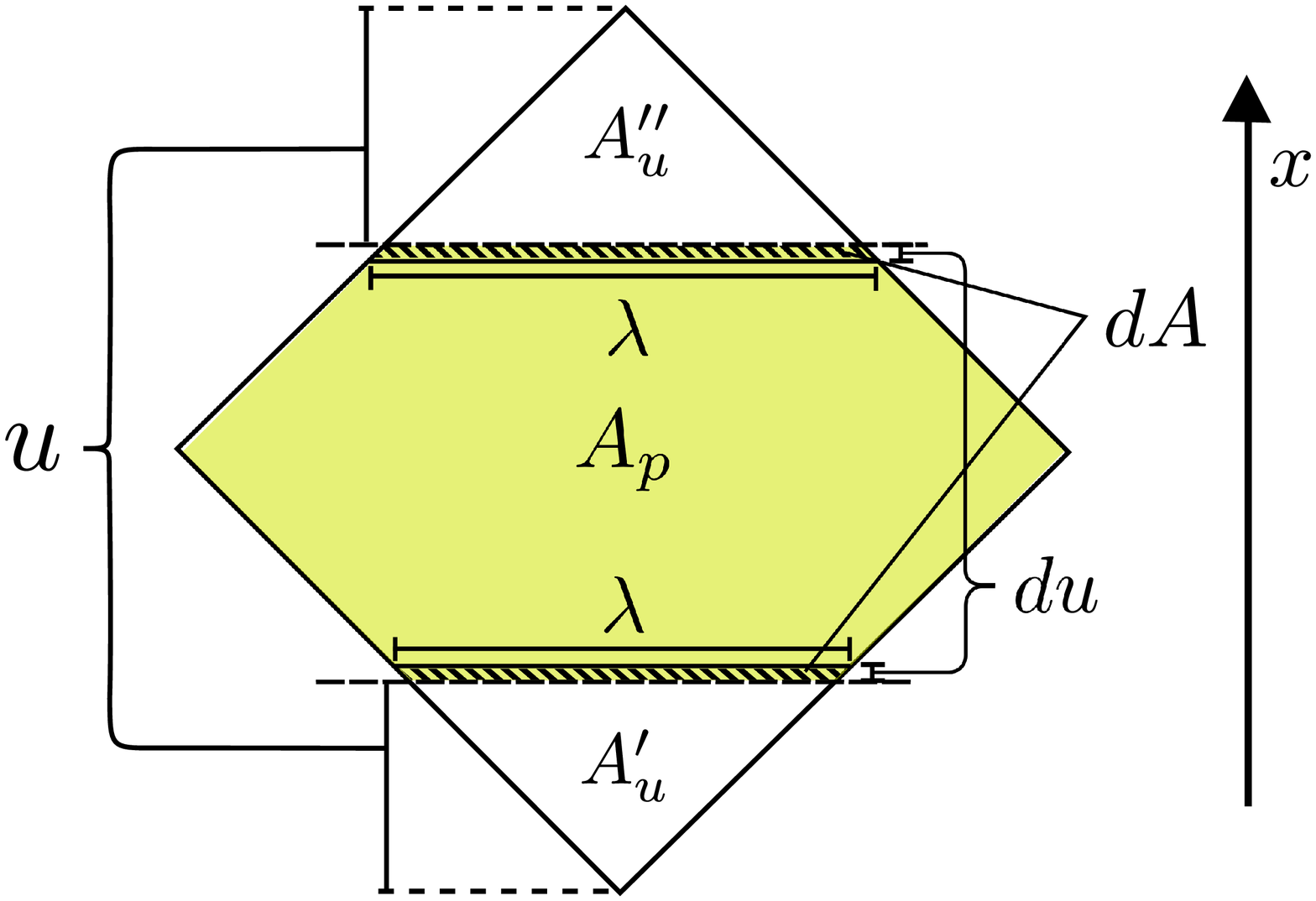}
    \caption{Scheme showing the infinitesimal unfolding of a packaged sheet. As the sheet is stretched of an element of length $dx$, the area element $dA$ (striped area) is unfolded. Note that the area element is equal to $\lambda du$.}
    \label{fig:calc_desd}
\end{figure}

The unfolding symmetry referred to in the main text is unidirectional, that is, the unfolding of the packaged phase grows along a single direction ($x$ axis), as can be seen in Figure 3-a. When stretching a crumpled sheet of an element of length $dx$, it is unfolded from a corresponding element of area $dA$, as shown in Fig.~\ref{fig:calc_desd}: the area element $dA$ is a rectangle with height $du = dx$ and length $\lambda$, whose area is equal to $\lambda du = \lambda dx$. The two unpacked parts have equal average lengths, so $u' = u''$, and their union forms an unpacked square sheet. This means that $\lambda = \delta$ is the length of the diagonal of a square, and then $dA = \delta dx$. Integrating $dA$ between, $0$ and $\delta_{f}$, we obtain

\begin{equation*}
    A_{f} = \int_{0}^{\delta_{f}} dA = \int_{0}^{\delta_{f}} \delta.dx = \frac{1}{2} \delta_{f}^2. 
\end{equation*}
When a sheet with size $L$ is unfolded up to its $\delta_{f}$, $A_{f} = L^2$, that is, the area of the unpacked phase, $A_{f}$, is equal to the total area of the sheet, $L^2$:

\begin{equation}\label{eq:L_prop_Dxf}\tag{S3}
    A_{f} = \frac{1}{2}\delta_{f}^2 \rightarrow L^2 = \frac{1}{2}\delta_{f}^2 \rightarrow L \propto \delta_{f}.
\end{equation}
The result of eq.~\ref{eq:L_prop_Dxf} shows that all crumpled sheets with $L > L_{c}$ have a dismantling dynamics described by eqs. 3 and 4.

\subsection{Extraction of peaks of stretching force}\label{apend:coex_fase}

To find a hierarchical behavior in the statistical distribution of peaks of force $F_p$ of the $F$-$\delta$ curves, we need define the regions of interest. In the first and second regions, the forces involved are locally transverse to the sheet domains; they generally connect separate domains on the sheet and are influenced by the hierarchy of the network of ridges. In the third region, the forces involved are internal to the sheet, this means that the corresponding slope for the curve $F$-$\delta$ is not related to the hierarchical interactions of the network of ridges and therefore, the peaks found in the third region were omitted from the analysis. The distribution of magnitude of peaks in the unpacking of crumpled sheets was reasonably well described by a lognormal distribution $P(x) \sim \exp[-(\ln(x) - \mu)^{2}/(2 \sigma^{2})]/(x \sigma)$ (In Fig. 4 the reader can find the distribution for sheets with size $L = 264$ mm).

We fitted the distributions of magnitude of peak in the unpacking of CS of sizes $L = \{$ 30, 55, 66, 77, 88, 147, 206, 264 and 305 $mm \}$, however, only the sheets with the three highest $L$ had a reasonable number of peaks to make a histogram. Probably the dynamometer used is not sensitive enough to record all fluctuations in the $F$-$\delta$ curve for sheets with small values of $L$. The standard deviation found for the unfolding of CS with sizes $L = 305, 264$ and $206$ mm were respectively $\sigma \approx 0.42, 0.45$ (Fig. 4) and $0.42$. Another distribution that gives a good fit is the Gamma distribution, $P(x) \sim x^{a-1}/[b^{a}\Gamma(a)]\exp(-x/b)$ with the shape parameter $a = 6.3, 4.9$ (Fig. 4) and $6.8$. However, the $\chi^{2}$ test showed that the log-normal distribution is better adjusted to the data.

%\bibliographystyle{ieeetr}
%\bibliography{name1}